\documentclass[12pt,a4paper,final]{article}

\usepackage[margin=22.4mm]{geometry}
\usepackage[utf8]{inputenc}  
\usepackage{xcolor,graphicx}
\usepackage[de-DE]{datetime2}  
\usepackage{amsmath,amsfonts,amssymb,mathrsfs,bm}
\usepackage{hyperref,relsize}

\usepackage{comment}
\newcommand{\BCOMM}{}

\usepackage{bef_alex} 

\newcommand{\qandq}{\quad\text{and}\quad}

\newcommand{\dotnabla}{{\bm\cdot}\nabla}
\newcommand{\BBR}{]\hspace{-0.3ex}]}
\newcommand{\BBL}{[\hspace{-0.3ex}[}

\def\FG{\bm}
\def\mafo{\mathrm}
\newcommand{\oti}{{\otimes}}
\renewcommand{\mdot}{{\bm:}}

\numberwithin{equation}{section}
\numberwithin{figure}{section}

\usepackage[notref,notcite]{showkeys}

\usepackage{upgreek}

\makeatletter
\renewcommand*\env@cases[1][1.2]{%
  \let\@ifnextchar\new@ifnextchar
  \left\lbrace
  \def\arraystretch{#1}%
  \array{@{\,}c@{\quad}l@{}}%
}
\makeatother

\newcommand{\sfq}{{\mathsf q}}
\newcommand{\sfw}{{\mathsf w}}
\newcommand{\sfS}{{\mathsf S}}
\newcommand{\sfx}{{\mathsf x}}
\newcommand{\Cauchy}{\bfSigma_\mathrm{Cauchy}}
\newcommand{\Fe}{\bfF_{\mathrm e}}
\newcommand{\Fp}{\bfF_{\mathrm p}}
\newcommand{\mfL}{\mathfrak L}
\newcommand{\mfT}{\mathfrak T}
\newcommand{\plt}[1]{\frac{\pl{#1}}{\pl t}}

\begin{document}

\title{An Eulerian formulation for dissipative materials\\
using Lie derivatives and GENERIC}

\author{Alexander Mielke
\thanks{Weierstrass-Institut f\"ur Angewandte Analysis und 
Stochastik, Mohrenstr.39, D-10117 Berlin, Germany}}

\date{27 February 2025} 
\maketitle

\centerline{\itshape In memory of  Wolfgang Dreyer} 

\begin{abstract}
  We recall the systematic formulation of Eulerian mechanics in terms of Lie
  derivatives along the vector field of the material points. Using
  the abstract properties of Lie derivatives we show that the transport via Lie
  derivatives generates in a natural way a Poisson structure on the chosen
  phase space.

The evolution equations for
thermo-viscoelastic-viscoplastic materials in the Eulerian setting is 
formulated in the abstract framework of GENERIC 
(General Equations for Non-Equilibrium Reversible Irreversible Coupling). 
The equations may not be new, but the systematic splitting between 
reversible Hamiltonian and dissipative effects allows us to see the equations
in a new light that is especially useful for future generalizing of the
system, e.g.\ for adding new effects like reactive species.
\end{abstract}




\section{Introduction}\label{sect-intro}

Motivated by the related work \cite{MieRou25GTMF} we reconsider the theory of
Lie derivatives for formulation dissipative continuum-mechanical systems 
in the Eulerian setting. For this we provide in Section \ref{se:LieDeriv} an
introduction to the theory of Lie derivatives, first in the differential
geometrical setting involving multi-linear forms and then for vectors,
co-vectors, and operators as they are used in continuum mechanics. The main
point here is to gain an understanding that the usage of such objects may have
different interpretations in differential geometry and hence need a suitable
Lie derivative. It is well-known that there are different Lie derivatives for
stress tensors (stress rates) but the issue is already relevant the distinction
of extensive and intensive field variables and 
for vectors and co-vectors such that as the momentum.

In Section \ref{se:GENERIC} we first recall the GENERIC framework, where  
the acronym GENERIC stands for {\it General Equation for Non-Equilibrium
  Reversible-Irreversible Coupling} and was introduced in
\cite{GrmOtt97DTCF12}.  However, this class of models originates in the
metriplectic theory developed in \cite{Morr84BFIC,Morr86PJHD}, cf.\ the survey
\cite{Morr09TBDO}.  Over the last decade, GENERIC has
proved to be a versatile modeling tool for various complex coupled models for
fluids and solids, see e.g.\ \cite{LeJoCa08UNET, Miel11FTDM, HutSve12TMFV,
  DuPeZi13GFVF, PaKlGr18MTDI, PaPeKl20HCM, BetSch19EMEC, Lasa21ATCN, PTPS22CNCF,
  ZaPeTh23GFRF} and the references therein.  We also refer to \cite{Miel15TCQM,
  KMMR19MMSQ} for applications in semiconductor and quantum devices. In 
\cite{MiPeZi24?DGSH} it is shown that a dissipative GENERIC system
can be rigorously derived from an infinite-dimensional (non-dissipative)
Hamiltonian systems.

In Section \ref{su:KinemEuler} we start to develop continuum
mechanics system at 
finite strain in a systematic way that is compatible with Lie derivatives and
GENERIC. In particular, we consider elastoplastic materials based on the
multiplicative split $\bfF = \Fe \Fp$ where $\bfF$ remains related with the
velocity field $\bfv$ through the kinematic relation 
\[
\pl_t \bfF = -\mfL^\mafo{ve}_\bfv \bfF := - \bfv\dotnabla \bfF +(\nabla\bfv) \bfF,
\]
which is indeed given by a suitable Lie derivative, see Lemma
\ref{le:KinemRela}. 

In addition to the momentum $\bfpi= \rho \bfv$ (with
$\rho=\rho_\mafo{ref}/\det\bfF$) and $\bfF$, we use the plastic distorting
$\Fp\in \mafo{GL}(\R^d)$ and a scalar thermodynamical field variable $\tau$ for
form the state variable $q=(\bfpi,\bfF,\Fp,\tau)$.  The variable $\tau$ can be
chosen quite general, e.g.\ as density of internal energy, enthalpy, or
entropy or as the temperature.

In Section \ref{su:EnerEntrDriv} we discuss how the total energy $\calE(q)$ and
the total entropy $ \calS(q)$ as main objects in the GENERIC framework generate
thermodynamical driving forces that generate the evolution equations
\[
\pl_t q = \bbJ(q)\rmD\calE(q) + \pl_{\zeta} \calR^*\big(q, \rmD\calS(q)\big) ,
\]
which include the balance laws of mechanics as well as the constitutive
relations.   Here $\bbJ$ is the Poisson operator, and $\calR^*$ is the
dual dissipation potential, see Section \ref{su:SetupGEN}. All the material
properties for the dissipative mechanisms are contained in $\calR^*$ and the 
mapping $\xi\mapsto \bfj:=\pl_\xi\calR^*(q,\xi)$ is called the abstract kinetic
relation between the vector $\xi$ of all thermodynamic driving forces and the 
vector $\bfj$ of all corresponding fluxes. 

The main theoretical result of this paper is Theorem \ref{th:JacobiLie} that
shows that there is a straightforward way for the construction of a
Poisson operator $q \mapsto \bbJ(q)$ for Eulerian mechanics based on the general
properties for Lie derivatives. It is usually difficult to show that $\bbJ$
satisfies the so-called Jacobi identity (see \eqref{eq:GEN.Jacobi}), but
exploiting the calculation rules for general Lie derivatives (in
particular the commutator rule \eqref{eq:LieCommutator}) they can be
established without too much efforts.

In Section \ref{se:GENERIC} we then follow the full program of the GENERIC
framework and derive our Eulerian model for thermo-visco-elastoplasticity  
using the multiplicative split $\bfF=\Fe\Fp$. One of the advantages of this
framework is that we are able to derive the reversible Hamiltonian part and the
irreversible dissipative parts independently, namely
\[
\pl_t q =\bfV_\mafo{Ham}(q) + \bfV_\mafo{diss}(q) \ \text{ with } 
\bfV_\mafo{Ham}(q):= \bbJ(q)\rmD\calE(q) 
\text{ and }  \bfV_\mafo{diss}(q):= \pl_{\zeta} \calR^*\big(q,
\rmD\calS(q)\big). 
\]
This allows us to study the different physical effects separately, thus
providing a better overview on the physical principles. 

Our final system can be written in terms of the variables $q=(\bfpi,\sfw)$
where $\bfpi=\rho\bfv$ is the momentum and $\sfw=(\bfF,\Fp,\tau) $, and it
takes the form
\begin{subequations}
 \label{eq:I.system}
  \begin{align}
  \label{eq:I.system-a}
  \pl_t \bfpi&= -\mfL_\bfv^\mafo{mo} \bfpi
    + \DIV\big(\bfSigma_\mafo{Cauchy}(\sfw) + \bbD_\mathrm{visc}(\sfw)
    \bfD(\bfv) \big), 
  \\[0.3em] 
  \label{eq:I.system-b}
  \pl_t \bfF &=-\mfL_\bfv^\mafo{ve} \bfF,
  \\[0.3em] 
  \label{eq:I.system-c}
  \pl_t\Fp &=-\mfL_\bfv^\mafo{in} \Fp + \Fp \bfL_\mafo{vi.pl} (\sfw),
  \\[0.3em] \label{eq:I.system-d}
  \pl_t \tau&=j^S_\mafo{Ham}(\sfw) +j^E_\mafo{diss}(\sfw)
   - \tdfrac{1}{\pl_\tau E(\sfw)} \DIV\!\Big( 
  \bbK_\mathrm{heat}(\sfq) \nabla\tdfrac1{\Theta(\sfw)}\Big).
  \end{align}
\end{subequations}
The first equation is the momentum balance including the Cauchy stress
tensor and the visco-elastic term with strain-rate tensor
$\bfD(\bfv)=\frac12\big(\nabla\bfv+(\nabla\bfv)^*\big)$. The second equation
states that the deformation gradient $\bfF$ is simply transported by a
(suitable) Lie derivative, and the third equation contains the visco-plastic
flow rule encoded by $\bfL_\mafo{vi.pl}$. The last equation is a scalar
thermodynamic equation which reduces (i) to the energy balance if $\tau$ is chosen
as the density of the internal energy, (ii) to the heat equation for the choice
$\tau=\theta =$\,temperature, or (iii) to the entropy (im)balance if we choose
$\tau$ as entropy density. We refer to Section \ref{su:FullGEE} for the
explanation of all symbols. 

At this stage we want to highlight that the  
first three equations clearly show the relevance of
the Lie derivative $\mfL_\bfv$, while in the fourth equation it will only
appear, if one chooses $\tau$ as an intensive or an extensive variable.
Here we have added the superscripts ``mo'', ``ve'', ``in'' for ``momentum'',
``vector'', and ``internal variable'' to the Lie
derivative $\mfL_\bfv$ to indicate that the Lie derivatives depend on the
tensorial nature of the variables $\bfpi$, $\bfF$, and $\Fp$, respectively. 

Section \ref{se:Conclusion} concludes the paper by a discussion of the
developed theory, in particular 
on the relevance of the proper usage of Lie derivatives and the GENERIC
framework.

\section{Lie derivatives}
\label{se:LieDeriv}

The theory of Lie derivatives is a well-known tool in differential
geometry as well as in 
Eulerian fluid and solid mechanics. It is relevant when tensors (functions,
vectors, forces, densities, etc.) are defined over a manifold and there is a
vector field $\bfv$ on this manifold, see e.g.\ \cite{Yano55TLDA},
\cite[Cha.\,1]{Chan83MTBH}, 
\cite[Sec.\,1.6]{MarHug94MFE}, and \cite[Sec.\,4.3]{MarRat99IMS}.  See also
\cite[footnote\,26]{MarHug94MFE} for historical remarks concerning the
first applications of Lie derivatives in
continuum mechanics. 

\subsection{Differential geometric approach}
\label{su:DiffGeo}
In our case the manifold is the Eulerian space $\R^d$ or a subset $\sfS$
thereof, called the body in Eulerian description. The tangent and cotangent
spaces are $\rmT_\sfx\sfS=\R^d$ and $\rmT_\sfx^*\sfS=\R^d_*$, respectively.
A vector field $\bfv$ on $\sfS$ means that $\bfv(\sfx)\in \rmT_\sfx \sfS$ for all
$\sfx \in \sfS$, and a
co-vector field $\bfalpha$ satisfies $ \bfalpha(\sfx) \in \rmT^*_\sfx\sfS$. In the
differential geometric setting, a tensor
$\bfA$ of type $(i_\circ,j_\circ)$ on $\sfS$, written as $\bfA \in
\mfT_{j_\circ}^{i_\circ}(\sfS)$, means that $\bfA(\sfx)$ is a multi-linear mapping
from $(\rmT_\sfx\sfS)^{i_\circ} \ti  (\rmT^*_\sfx\sfS)^{j_\circ} $, i.e.\ linear in
each of the $i_\circ{+}j_\circ$ arguments. By the dual pairing  
\[
(\bfv, \bfalpha) \mapsto \langle \bfalpha,\bfv\rangle_\sfS: x \mapsto \langle
\bfalpha(\sfx),\bfv(\sfx)\rangle_{\rmT_\sfx\sfS}
\]
we can identify $\bfv$ with a tensor in $\mfT^0_1(\sfS)$ and  $\bfalpha$ with
a tensor in $\mfT^1_0(\sfS)$. The anti-symmetric tensors in
$\mfT^0_{j_\circ}(\sfS)$ are the differentials forms denoted by the set
$\Lambda_{j_0} (\sfS)$. 

For a general vector field $\bfv$ and a tensor field $\bfA$, the Lie derivative
is defined by taking the derivative of $\bfA$ along the flow of 
$\bfv$. Throughout this section we assume that all objects like $\bfv$ and
$\bfA$ are independent of time. First observe, that $\bfv$ defines a flow map
$s\mapsto \bfPsi(s,\cdot):\sfS \to \R^d$ such that the following ODE is satisfied:
\[
\frac{\rmd}{\rmd s} \bfPsi(s,\sfx) = \bfv(\bfPsi(s,\sfx)) \text{ for } s\in
{]{-}\delta,\delta[}, \qquad \bfPsi(0,\sfx) = \sfx. 
\]
The Lie derivative of $\bfA$ along $\bfv$ is now defined via 
\begin{equation}
  \label{eq:def.LieDeriv}
  \mfL_\bfv \bfA = \big( \frac{\rmd}{\rmd s} \bfPsi(s,\cdot)^* 
  \bfA\big)\Big|_{s=0},
\end{equation}
where the pull-back $\bfPhi^*\bfA$ of $\bfA$ by $\bfPhi$ is defined via
\[
\bfPhi^*\bfA(x)\big[\bfv_1,..,\bfv_{i_\circ},\bfalpha_1,..,\bfalpha_{j_\circ}\big]
= \bfA (\bfPhi(x))\big[\rmD\bfPhi\bfv_1,.., \rmD\bfPhi\bfv_{i_\circ};
\big( \rmD\bfPhi\big)^{-*} \bfalpha_1,.., \big(\rmD\bfPhi\big)^{-*}
\bfalpha_{j_\circ} \big] ,
\]
where $\rmD\bfPhi$ is evaluated at $x\in \calB$, see Appendix
\ref{se:FormDiffGeo}.   In particular, $\mfL_\bfv$ is
again a tensor of the same order as $\bfA$.

Because we have $\bfPsi(s,\sfx) = \sfx +s \bfv(\sfx)+ O(s^2)$  
we find $\rmD \bfPsi(s,\sfx)=\bfI+ s \nabla \bfv(\sfx) +
O(s^2)$ and conclude 
\begin{equation}
  \label{eq:LieDerivFormula}
\begin{aligned}
  \mfL_\bfv \bfA\big[ \bfv_1,..,\bfv_{i_\circ}; 
 \bfalpha_1,..,\bfalpha_{j_\circ}\big] 
&  =    \big(\bfv\dotnabla \bfA\big)
\big[ \bfv_1,...,\bfv_{i_\circ}; 
 \bfalpha_1,...,\bfalpha_{j_\circ}\big]
\\
&\quad + \sum_{i=1}^{i_\circ} \bfA\big[ \bfv_1,..., (\nabla
\bfv)\bfv_i,...,\bfv_{i_\circ};  \bfalpha_1,...,\bfalpha_{j_\circ}\big]
\\&\quad 
-\sum_{j=1}^{j_\circ} \bfA\big[ \bfv_1,...,\bfv_{i_\circ};
 \bfalpha_1,..., \big(\nabla \bfv)^*\bfalpha_j,..., \bfalpha_{j_\circ}\big].
\end{aligned}
\end{equation}

One of the fundamental properties of Lie derivatives is the commutator relation
\begin{equation}
  \label{eq:LieCommutator}
  \mfL_\bfv\big( \mfL_\bfw \bfT\big) - \mfL_\bfw \big(
\mfL_\bfv \bfT\big) = \mfL_{\BBL \bfv,\bfw\BBR } \bfT,
\end{equation}
where the commutator between vector fields is given by 
\[
\BBL \bfv,\bfw\BBR := (\nabla \bfw) \bfv -(\nabla \bfv) \bfw =
  (\bfv\dotnabla )\bfw -   (\bfw\dotnabla ) \bfv= \mfL_\bfv \bfw = - \mfL_\bfw \bfv .
\]
It is easily checked that we have the Jacobi identity for vector fields, viz.  
\begin{equation}
  \label{eq:JacoId.VectFie}
  \big[\!\big[ \bfv_1,\BBL \bfv_2,\bfv_3\BBR  \big]\!\big] + 
  \big[\!\big[ \bfv_2,\BBL \bfv_3,\bfv_1\BBR  \big]\!\big] + 
  \big[\!\big[ \bfv_3,\BBL \bfv_1,\bfv_2\BBR  \big]\!\big]  =0 \ \text{ for all } 
\bfv_1,\bfv_2,\bfv_3 \in \mfT^1(\sfS). 
\end{equation}
The identities \eqref{eq:LieCommutator} and \eqref{eq:JacoId.VectFie} will be
extremely useful for showing that 
the the operator $\bbJ$ to be constructed satisfies the Jacobi identity, and
hence is a Poisson structure. It is not easy to find explicit statements of 
\eqref{eq:LieCommutator} in the literature, 
but it is an easy consequence of its validity for functions, vectors, and
co-vectors ($1$-forms) and of the well-known derivation rule
$\mfL_\bfv (\bfT{\otimes}\bfS) = (\mfL_\bfv \bfT){\otimes}\bfS + \bfT
{\otimes}(\mfL_\bfv \bfS)$ for tensor products by doing induction over the rank
of the tensors (see e.g.\ condition ``(DO1)'' in \cite[Ch.\,2]{AbrMar78FM}).

Here the tensor product of $\bfA \in \mfT^{i_\circ}_{j_\circ}(\sfS)$ and
$\bfA \in \mfT^{n_\circ}_{m_\circ}(\sfS)$ is the tensor $\bfA\oti \bfB \in
\mfT^{i_\circ+n_\circ}_{j_\circ+m_\circ}(\sfS) $ defined via 
\[
\begin{aligned}
\bfA\oti \bfB [\bfv_1,\ldots,\bfv_{i_\circ+n_\circ};\bfalpha_1,\ldots,
\bfalpha_{j_\circ+m_\circ}] &:= \bfA[\bfv_1,\ldots,\bfv_{i_\circ};\bfalpha_1, 
  \ldots, \bfalpha_{j_\circ}]\\
&\qquad \cdot \bfB [\bfv_{i_\circ+1},\ldots,\bfv_{i_\circ+n_\circ}; 
 \bfalpha_{j_\circ+1},\ldots, \bfalpha_{j_\circ+m_\circ}] .
\end{aligned}
\]
Inner products can be defined by tensor products and subsequent contraction of
indices. For $\bfA \in \mfT^{i_\circ}_{j_\circ}(\sfS)$  and $n,m\in \N$ with 
$1\leq n\leq i_\circ$ and $1\leq m\leq j_\circ$ we can contract the $n$th
vector slot with the $m$ co-vector slot to obtain $\mathsf C^n_m \bfA \in
\mfT^{i_\circ-1}_{j_\circ-1}(\sfS)$ defined via  
\begin{align*}
\mathsf C^n_m \bfA &\big[\bfv_1,\ldots,\bfv_{i_\circ-1}; \bfalpha_1,\ldots
,\bfalpha_{j_\circ-1}\big] 
\\
& :=  \sum_{k=1}^d \bfA
\big[\bfv_1,..,\bfv_{k-1},\bfe_k,\bfv_k,..,\bfv_{i_\circ-1}; \bfalpha_1, 
 ..,\bfalpha_{m-1},\bfepsilon_k,\bfalpha_m,\ldots ,\bfalpha_{j_\circ-1}\big] ,
\end{align*}
where $\{\bfe_1,\ldots,\bfe_d\}$ is an arbitrary basis in $\rmT_\sfx\sfS$ and
$\bfepsilon_k \in \rmT_\sfx^*\sfS$ satisfy $\langle \bfepsilon_k, \bfe_l\rangle =
\delta_{k,l}$ (the Kronecker symbol) for $k,l\in \{1,\ldots,d\}$. 

\subsection{Differential forms}
\label{su:DifferForms}

The special inner product with a vector field $\bfw$ is denoted by $\bfi_w:
\mfT^{i_\circ+1}_{j_\circ}(\sfS) \to \mfT^{i_\circ}_{j_\circ}(\sfS)$ and is
given by 
\[
\bfi_\bfw \bfA\big[\bfv_1,..,\bfv_{i_\circ};\bfalpha_1,..,\bfalpha_{j_\circ}\big]
=\bfA\big[\bfw,\bfv_1,..,\bfv_{i_\circ};\bfalpha_1,..,\bfalpha_{j_\circ}\big] ,  
\]
i.e.\ by simple insertion into the first vector slot. Identifying $\bfw$ with
the tensor  $\bfB[\bfalpha]= \langle \bfalpha,\bfw\rangle$ we have the relation 
$\bfi_\bfw \bfA = \mathsf C^1_1\big(\bfA\oti \bfB\big)$. It is not difficult to
see that Lie derivatives commute with contraction, hence we also have the
product rule for inner products:
\[
\mfL_\bfv \mathsf C^n_m\bfA = C^n_m \mfL_\bfv \bfA \quad \mfL_\bfv \big(
\bfi_\bfw \bfA) = \bfi_\bfw (\mfL_\bfv \bfA)  + \bfi_{\mfL_\bfv \bfw} \bfA.
\]

An important subclass of tensor is given by the differential forms that 
appear as $k$-forms for $k\in \{0,\ldots, d\}$. They are given as tensors
$\bfbeta \in \Lambda_m(\bfS) \subset \mfT^0_k(\sfS)$, where $\Lambda_m$
denotes the subset of tensors that are anti-symmetric, i.e.\ interchanging any
two arguments changes the result by a factor $(-1)$. One can define the
differential $\bfd: \Lambda_k(\bfS) \to \Lambda_{k+1}(\sfS)$ and has
$\bfi_\bfw: \Lambda_{k+1}(\bfS)\to \Lambda_k(\bfS) $.  For differential forms
$\bfbeta$ one has the identities
\[
\mfL_\bfv \bfbeta = \bfi_\bfv \bfd \bfbeta + \bfd(\bfi_\bfv \bfbeta) \qandq 
\mfL_\bfv (\bfd \bfbeta) = \bfd\big( \mfL_\bfv \bfbeta\big),
\]
where the first one is called ``Cartan's magic formula'' (also known as
Cartan's first structural equation) and the second follows easily
from the first when using $\bfd\bfd\bfgamma=0$ for all differential forms
$\bfgamma$.  

The wedge product $\wedge$ maps $\Lambda_{k}(\sfS)\ti \Lambda_l(\sfS)$ into
$\Lambda_{k+l}(\sfS)$ by $\bfbeta \wedge \bfgamma = \mafo{anti}(\bfbeta\oti
\bfgamma)$ and satisfies
\[
\mfL_\bfv (\bfbeta \wedge \bfgamma) = \big(\mfL_\bfv \bfbeta\big) \wedge \bfgamma
+ \bfbeta \wedge \big(\mfL_\bfv \bfgamma \big) .
\] 

\subsection{Tensors with values in a linear space}
\label{su:U.valuedTensors}

The above theory can easily generalized to the case that the tensors $\bfA(x)$ do
not map into $\R$ but into a general linear space $\bfU_\sfx$, i.e.\ 
\[
\bfA(x): (\rmT_\sfx\sfS)^{i_\circ} \ti (\rmT_\sfx^*\sfS)^{j_\circ} \to \bfU_\sfx
\]
is still multi-linear in $(\bfv_1,\ldots,\bfv_{i_\circ}
;\bfalpha_1,\ldots,\bfalpha_{j_\circ})$. We will then speak of
$\bfU$-valued tensors and remark that $\bfU_\sfx$ has to be independent of
$\rmT_\sfx\sfS$ and $\rmT_\sfx^*\sfS$, which means that the vector-field $\bfv$ is
not moving the points in $\bfU_\sfx$.

\subsection{Lie derivatives of vectors, operators, and volume forms}
\label{su:LieDerVecOper}

A proper usage of differential geometric concepts in continuum mechanics
is not standard, but we refer to \cite{Sege23FGCM} for a careful and detailed
approach in this direction. Traditionally, it is more common to use vector,
co-vector and matrices, called tensors there. To distinguish these notions we
use the name ``operator'' for matrices acting as linear mappings between linear
spaces.  

Even for scalar-valued functions the Lie derivatives is not simple, because
functions in continuum mechanics occur in different forms. First one has
\emph{intensive} and \emph{extensive} functions, which have the differential
geometric interpretation as $0$-forms (i.e.\ $f\in \Lambda_0(\sfS)$) and
volume or $d$-form (i.e.\ $\bfrho \in \Lambda_d(\sfS)$), respectively. In the
latter case identify a function $\rho$ with the volume form
$\bfrho[\bfv_1,\ldots,\bfv_d]= \rho\,\det\big(\bfv_1|..|\bfv_d\big)$, and we obtain 
\begin{equation}
  \label{eq:LieD.0.d.form}
  \mfL_\bfv f= \bfv{\cdot} \nabla f \qandq  \mfL_\bfv \rho  = \DIV( \rho \bfv). 
\end{equation}
In continuum mechanics the typical extensive variables are mass density $\rho$,
internal-energy density $e$, entropy density $s$, or any one-homogeneous
function of those. The typical intensive variables are concentrations,
temperature $\theta = \rmd e/\rmd s$, velocity, pressure, or chemical
potentials. For general functions, not falling into the classes of intensive or
extensive functions, the Lie derivative has to be obtained by representing it
via intensive and extensive variables and then applying the chain rule and the
appropriate Lie derivatives.

A vector field $\bfw$ and a co-vector field $\bfalpha$ can be
identified with tensors $\bfA_\bfw \in \mfT^0_1(\sfS)$ and $\bfB_\bfbeta\in
\mfT^1_0(\sfS)$, respectively, by 
\[
\bfA_\bfw[\bfalpha] = \langle \bfalpha,\bfw\rangle \qandq  
\bfB_\bfbeta[\bfv] = \langle \bfbeta,\bfv\rangle .
\]
Using this, we can derive the form of the Lie derivatives from the tensor
Lie derivatives, namely
\begin{equation}
  \label{eq:LieD.vec.covec}
  \mfL_\bfv \bfw = \bfv\dotnabla \bfw -(\nabla\bfv)\bfw=\BBL \bfv, \bfw\BBR  \qandq 
\mfL_\bfv \bfbeta = \bfv\dotnabla\bfbeta +(\nabla\bfv)^* \bfbeta.
\end{equation} 
Moreover, in continuum mechanics the momentum vector $\bfpi=\rho I_\rmR\bfv$ (where
$I_\rmR:\rmT_\sfx\sfS \to \rmT_\sfx^*\sfS)$ is the Riesz isomorphism) is a
co-vector but its correct interpretation is as a $(d{-}1)$-form (by the Hodge
star operator mapping $\Lambda_k(\sfS) $ into $\Lambda_{d-k}(\sfS)$), namely 
\begin{subequations}
\label{eq:Momentum}
\begin{equation}
\label{eq:moment.d-1form}
\bfpi = \rho I_\rmR \bfv = \bfi_\bfv \bfrho .
\end{equation}
This leads to the Lie derivative 
\begin{equation}
\label{eq:LieDer.moment}
\mfL_\bfv \bfpi = \DIV(\bfpi\oti \bfv) + (\nabla \bfv)^* \bfpi,
\end{equation}
\end{subequations}
see \cite[Prop.\,4.2]{MieRou25GTMF}. 
A ``vector'' may also be considered as a tensor with values in the linear
$\R^d$ which is not connected to $\rmT_\sfx\sfS$. Such a case is occurs in
\cite[Eqn.\,(2.3)]{BNSS24?VAMC}, where the magnetization vector $\bfM(t,\sfx)\in
\R^3$ has the Lie derivative $\mfL^\mafo{mag}_\bfv \bfM= \bfv\dotnabla M +
(\DIV \bfv) \bfM= \DIV\big( \bfM\oti \bfv\big)$. The proper tensorial
interpretation of ``flux fields'' is reported in
\cite[p.\,71]{Sege23FGCM}:  {\small ``A flux field, such as the heat flux
field, should be regarded as a 2-form in the three-dimensional body rather than as a
vector field. As a 2-form, a heat flux field may be considered independently of the
configuration of the body in space.''} 
\medskip

We have four types of operators, namely
\[
\bbB:\rmT_\sfx \sfS\to \rmT_\sfx\sfS, \quad 
\bbC:\rmT_\sfx \sfS\to \rmT_\sfx^*\sfS, \quad 
\bbD:\rmT_\sfx^* \sfS\to \rmT_\sfx^*\sfS, \quad 
\bbE:\rmT_\sfx^* \sfS\to \rmT_\sfx\sfS.
\]
They can be identified with tensors $\bfB\in \mfT_1^1$, $\bfC\in \mfT^2_0$,
$\bfD \in \mfT_1^1$, and $\bfE\in \mfT_2^0$ via
\[
\bfB[\bfv,\bfalpha]=\langle \bfalpha, \bbB\bfv\rangle, \quad
\bfC[\bfv_1,\bfv_2]=\langle \bbC\bfv_1,\bfv_2\rangle, \quad
\bfD[\bfv,\bfalpha]=\langle \bbD \bfalpha, \bfv\rangle, \quad
\bfE[\bfalpha_1,\bfalpha_2]=\langle \bfalpha_1, \bbE\bfalpha_2\rangle.
\]
Thus, the corresponding Lie derivatives can be calculated by those for $\bfB$
to $\bfE$ and we obtain  
\begin{equation}
  \label{eq:TensorRates}
\begin{aligned}
&\mfL_\bfv \bbB=\bfv \dotnabla \bbB + \bbB (\nabla \bfv) -(\nabla\bfv)\bbB,  
& \mfL_\bfv\bbC=\bfv\dotnabla\bbC+\bbC(\nabla\bfv) + (\nabla\bfv)^*\bbC,
\\
&\mfL_\bfv\bbD=\bfv\dotnabla\bbC -\bbD (\nabla\bfv)^*+(\nabla\bfv)^*\bbD,
&\mfL_\bfv\bbE=\bfv\dotnabla\bbE- \bbE(\nabla\bfv)^*- (\nabla\bfv)\bbE.
\end{aligned}
\end{equation}
Note that only $\bbC$ and $\bbE$ can be symmetric tensors and that in this
case the two signs of the second and third terms are the same and one times we
have $\nabla\bfv$ and the other time $(\nabla\bfv)^*$. See also
\cite[Ch.\,1\;Box\,6.1]{MarHug94MFE}. 

As an example consider the Euclidean tensor $\bfC_\mafo{Euc} \in \mfT^2_0$ with
$\bfC_\mafo{Euc} [\bfv_1,\bfv_2] = (\bfv_1|\bfv_2)_\mafo{Euc}$ (the Euclidean
scalar product). The associated operator
$\bbC_\mafo{Euc}=\bbI_\mafo{Riesz}:\rmT_\sfx\R^3\to 
\rmT^*_\sfx\R^3$ is the Riesz isomorphism. The Lie derivative provides twice the
classical strain-rate tensor 
\begin{equation}
  \label{eq:def.bfD.bfv}
    \mfL_\bfv \bbC_\mafo{Euc} = 2 \,\bfD(\bfv) \quad \text{with } 
\bfD(\bfv):=\frac12 \big( \bbI_\mafo{Riesz}\nabla\bfv +
(\nabla\bfv)^*\bbI_\mafo{Riesz}\big) \in
\mafo{Lin_{sym}}\big(\R^3;(\R^3)^*\big).  
\end{equation}

There are various objective stress rates which arise from the fact that there
are various stress tensor which have different tensorial properties. 
Similar to the construction of the momentum as a $(d{-}1)$ form
\eqref{eq:Momentum} one also has to consider the Cauchy stress tensors
$\Cauchy$ as an extensive version, and its Lie derivative is the so-called
Truesdell stress rate 
\[
\mfL_\bfv \Cauchy =  \bfv \dotnabla \Cauchy + (\DIV\bfv)  \Cauchy -
\Cauchy (\nabla\bfv)^* - (\nabla\bfv)\Cauchy . 
\] 
However, the Kirchhoff stress tensor 
$\bfT_\mafo{Kir} = \det \bfF\, \Cauchy= \frac{\rho_\mafo{ref}}{\rho} \Cauchy $
(see Section \ref{su:KinemEuler} for the last identity) has ``intensive''
properties, because it is the quotient of two extensive objects. Thus,
$\bfT_\mafo{Kir} $ has to be interpreted as element of $\mfT^0_2$ like
$\bfE$ in \eqref{eq:TensorRates}, and its Lie derivative is the upper convected
or Oldroyd stress rate
\[
\mfL_\bfv \bfT_\mafo{Kir}=  \bfv \dotnabla \bfT_\mafo{Kir} -
\bfT_\mafo{Kir}(\nabla\bfv)^* - (\nabla\bfv)\bfT_\mafo{Kir} .
\]
Here, we do not dwell on this subject any further, but refer to 
\cite[Ch.\,1\;Box\,6.1]{MarHug94MFE} for this and to \cite{Fial08GFDL} for a
way to write the Zaremba-Jaumann derivative as a Lie derivative.

\section{GENERIC structures and  Eulerian mechanics}
\label{se:GENERIC}

The acronym GENERIC stands for 
\begin{quote}
{\it General Equation for Non-Equilibrium
  Reversible-Irreversible Coupling} 
\end{quote}
and was introduced in
\cite{GrmOtt97DTCF12},  but the class of models appears metriplectic systems
already in \cite{Morr84BFIC,Morr86PJHD}, cf.\ the survey
\cite{Morr09TBDO}.  Over the last decade, the {\smaller GENERIC} framework has
proved to be a versatile modeling tool for various complex coupled models for
fluids and solids, see e.g.\ \cite{LeJoCa08UNET, Miel11FTDM, HutSve12TMFV,
  DuPeZi13GFVF, PaKlGr18MTDI,BetSch19EMEC, Lasa21ATCN, PTPS22CNCF,
  ZaPeTh23GFRF, MieRou25GTMF} and the references therein.

\subsection{Setup of GENERIC}
\label{su:SetupGEN}

We consider states $q$ in a state space $\bfQ$ which is either a flat space or a
smooth manifold. A GENERIC system is a quintuple
$(\bfQ,\calE,\calS,\bbJ,\bbK)$, 
where the energy $\calE$ and the entropy $\calS$ are differentiable functions
on $\bfQ$ with differentials $\rmD\calE(q)$ and $\rmD\calS(q)$ lie
$\rmT_q\bfQ$. The geometric structures are the Poisson operator $\bbJ$ for
Hamiltonian dynamics and a (dual) dissipation potential $\calR^*$, where
$\bbJ(q)$ maps $\rmT^*\bfQ $ to 
$\rmT \bfQ$ and $\calR^*(q,\cdot):\rmT^*_q Q\to [0,\infty]$ is a convex
functional with $\calR^*(q,0)=0$. In many cases $\calR^*(q,\cdot)$ is
a quadratic functional given in terms of a symmetric, positive
semi-definite operator $\bbK(q)$, namely
$\calR^*(q,\#xi)=\frac12\langle \xi, \bbK(q)\xi\rangle$. Such operators
$\bbK(q)$ are often called ``Onsager operators'' because of Onsager's
fundamental reciprocal relations in \cite{Onsa31RRIP}, earning him the Nobel
Prize in Chemistry in 1968.

The evolution equation then takes the form 
\[
\plt q = \bbJ(q) \rmD\calE(q) + \pl_\xi\calR^*\big(q, \rmD \calS(q) \big),
\]
where $\pl_\xi\calR^*(q,\cdot)$ denotes the convex subdifferential of
$\calR^*(q,\cdot)$, which is possibly set-valued (e.g.\ in plasticity).

The Poisson operator is defined by being skew-symmetric and satisfying the
Jacobi identity, i.e.\ 
\begin{equation}
  \label{eq:GEN.Jacobi}
  \big\langle \zeta_1, \rmD\bbJ(q)[\bbJ(q)\zeta_2] \zeta_3\big\rangle +
\text{cycl.\,perm.}\equiv 0\ \ \text{ for all }\ \zeta_1,\zeta_2,\zeta_3 \in \rmT^*_q\bfQ.
\end{equation}

The main condition for GENERIC are the so-called \emph{non-interaction
  conditions}, namely 
\begin{equation}
\label{eq:GENERIC.NIC}
\text{(a) } \ 
\bbJ(q) \rmD\calS(q)\equiv 0 \qquad \text{and\qquad (b) }\ 
 \calR^*(q,\lambda \rmD\calE(q))\equiv 0 \  \text{ for all
}\lambda \in \R.  
\end{equation}
By convexity, the latter condition implies $\calR^*(q,\zeta{+}\lambda
\rmD\calE(q)) = \calR^*(q,\zeta)$ for all $(q,\zeta) \in \rmT^*\bfQ$ and $\lambda
\in \R$, and as a consequence we have $\big\langle
\rmD\calE(q)),\pl_\xi\calR^*(q,\xi) \big\rangle =0$ for all $\xi$. 

Using the chain rule, a simple consequence of Condition
(\ref{eq:GENERIC.NIC}.b) (and $\bbJ=-\bbJ^*$) is the conservation of energy
along solutions, i.e.\ $\frac\rmd{\rmd t} \calE(q(t))=0$.  Condition
(\ref{eq:GENERIC.NIC}.a) (together with convexity of $\calR^*(q,\cdot)$)
implies entropy increase:
$\frac\rmd{\rmd t} \calS(q(t))=\big\langle \rmD\calS(q),
\pl_\xi\calR^*\big(q,\rmD\calS(q)\big) \big\rangle$ $\geq 0$, i.e.\ the
second law of thermodynamics is automatically satisfied for GENERIC
systems.

However, we emphasize that \eqref{eq:GENERIC.NIC} is much stronger that energy
conservation and entropy increase.  We refer to \cite{Otti05BET} and
\cite[Sec.\,2.2]{Miel11FTDM} for further properties of GENERIC systems, in
particular concerning additional conservation laws and the maximum entropy
principle providing thermal equilibrium states when maximizing $\calS(q)$
subject to the conserved quantities. We also refer to \cite{MGKO00SECG} for a
discussion on the differences between GENERIC and other thermodynamical
approaches.

An efficient way of constructing GENERIC systems is using the ``special form'' of
GENERIC systems as described in \cite[Sec.\,2.3.2]{Otti05BET} (following
\cite{Edwa98ASDG}) and \cite[Sec.\,2.4+4.3]{Miel11FTDM}.  
Having a suitable ``simpler'' Poisson structure $\bbJ_\mafo{simple}$ and a
``simpler'' dual dissipation potential $\calR^*_\mafo{simple}$, one
can define more complex structures via 
\begin{subequations}
  \label{eq:SpecGen.all}
\begin{equation}
  \label{eq:SpecGEN.J.R}
  \bbJ(q)=M_\calS(q) \bbJ_\mafo{simple}(\Phi(q))M_\calS(q)^* \quad \text{and} \quad
  \calR^*(q,\xi)= \calR^*_\mafo{simple}\big(q,M_\calE(q)^*\xi\big) 
\end{equation}
where the following additional conditions have to be met:
\begin{align}
&  \label{eq:SpecGEN.MS.ME} M_\calS(q)^*\rmD\calS(q)=f_*, \quad \bbJ_\mafo{simple}(q)f_*=0, \quad 
 M_\calS(q) = \big(\rmD \Phi(q)\big)^{-1}, \\
&  \label{eq:SpecGEN.si.f*g*}
  M_\calE(q)^*\rmD\calE(q)=g_*, \quad \calR^*_\mafo{simple}(q,\lambda g_*)=0
  \text{ for all }\lambda\in \R,
\end{align}
\end{subequations}
where $f_*$ and $g_*$ are fixed vectors.

The typical application occurs in continuum systems where $q=(w,\tau)$ and
$\tau$ is a scalar thermodynamical variable such as temperature $\theta$, internal
energy density $u$, or the entropy density $s$. Then, we set
$\Phi(w,\tau)=(w,S(w,\tau))^\top$ and obtain  
\[
M_\calS(w,\tau)=\bma{@{}cc@{}}I&0\\ \pl_w S(w,\tau)& \pl_\tau S(w,\tau)\ema^{-1} =
\bma{@{}cc@{}} I & 0\\ 
 -\frac{\langle\pl_w S(w,\tau),\Box\rangle}{\pl_\tau S(w,\tau)} &
\frac1{\pl_\tau S(w,\tau)}\ema.  
\]
Here and below the symbol ``$\Box$'' indicates at which position the
corresponding argument from matrix multiplication has to placed. In an
example this means 
\[
  \bma{cc} a\,\nabla \Box & \ee^{\Box \tau}V \\ \Box \nabla \phi & 
  M\,\nabla \Box \ema \binom\xi\eta = \binom{a \nabla \xi  + \ \  \ee^{\eta
      \tau}V\ \ } { \xi \nabla \phi + M\,\nabla \eta }.  
\]
With  the similar construction for $\calE$ we obtain 
\begin{align*}
&M_\calS(w,\tau)^*=
\bma{cc@{}} I& -\frac{\Box}{\pl_\tau S(w,\tau)}\,\pl_wS(w,\tau) \\ 0 &
\frac1{\pl_\tau S(w,\tau)} \ema = \bma{@{}cc@{}}I&0\\ 0&1\ema
+ \binom{-\pl_wS}{1{-}\pl_\tau S} \oti \binom{0}{\frac1{\pl_\tau S}} 
 \text{ and } 
\\
&
M_\calE(w,\tau)^*=
\bma{cc@{}} I& -\frac{\Box}{\pl_\tau E(w,\tau)}\,\pl_wE(w,\tau) \\ 0 &
\frac1{\pl_\tau E(w,\tau)} \ema = \bma{@{}cc@{}}I&0\\ 0&1\ema
+ \binom{-\pl_wE}{1{-}\pl_\tau E} \oti \binom{0}{\frac1{\pl_\tau E}} \!.
\end{align*}
Clearly, we have \eqref{eq:SpecGEN.MS.ME} with $f_*=g_*=(0,1)^\top$. The
remaining relation \eqref{eq:SpecGEN.si.f*g*} will follow from the classical
balance equations via $\nabla 1\equiv 0$. 

Looking at $M_\calS$ the best choice for $\tau$ is the entropy density $\tau
=s$ such that $S(w,s)=s$, because this gives $M_\calS(w,s)^*=\bbI$. However, this
choice is not optimal for $M_\calE$. In engineering often $\tau=\theta$ is
chosen, which makes both matrices $M_\calS$ and $M_\calE$ nontrivial.

\subsection{Eulerian thermo-elastoplasticity and its kinematics}
\label{su:KinemEuler}

In Lagrangian mechanics a body is described on a reference domain $\rmM
\subset \R^d$ with material points $\rmx \in \rmM$.  The time-dependent
deformation is denoted by $\sfx = \bfy(t,\rmx)$, where $\sfx \in \sfS$ is the
spatial coordinate. Denoting the inverse  mapping (also called return mapping)
by $\rmx=\rmY(t,\sfx)$
the Eulerian velocity $\bfv$ and the Eulerian deformation gradient $\bfF$ are
given by 
\[
\bfv(t,\sfx) = \frac{\pl}{\pl t} \bfy\big(t,\rmY(t,\sfx)\big) \quad \text{and} \quad 
\bfF(t,\sfx)= \nabla_{\!\rmx} \bfy \big( t, \rmY(t,\sfx) \big).
\]
Note that $\bfv$ is a vector field on $\sfS\subset \R^d$, whereas $\bfF(t,x)$
is a two-point tensor mapping $\rmT_{\rmY(t,\sfx)}\rmM$ to $\rmT_\sfx \sfS$. Thus,
we have to understand $\bfF$ as a $\bfU$-valued 1-form on $\calS$, where
$\bfU_\sfx=\rmT_{\rmY(t,\sfx)}\rmM$. The important kinematic relation is that 
$\bfF$ is directly transported by its Lie derivative. The same happens to the
Eulerian density $\rho(t,\sfx)= \rho_\mafo{ref}/\det\big(\bfF(t,\sfx)\big)$,
where the referential density $\rho_\mafo{ref}$ is assumed to be constant.

These result are well-known, and we state them to  highlight their Lie
derivative aspect.  

\begin{lemma}[Kinematic relation]
\label{le:KinemRela}
Considering $\bfv$, $\bfF$, and $\rho$ as above we have the relations (see
\eqref{eq:LieD.0.d.form} and \eqref{eq:LieD.vec.covec}) 
\[
\pl_t \rho= - \mfL_\bfv \rho=- \DIV(\rho\bfv) \quad
\text{and} \quad \pl_t \bfF = -
\mfL_\bfv \bfF= - \bfv\dotnabla \bfF +(\nabla\bfv)\bfF. 
\]
\end{lemma}     

In finite-strain elastoplasticity, the deformation tensor $\bfF$ is
multiplicatively decomposed by the so-called  Kr\"oner-Lee-Liu form 
(cf.\ \cite{Kron60AKVE,LeeLiu67FSEP})
\begin{equation}
  \label{eq:KronerLeeLiu}
  \bfF(t,\sfx) =\Fe(t,\sfx) \!\;\Fp(t,\sfx),
\end{equation}
where $\Fp(t,\sfx)$ maps $\rmT_{\rmY(t,\sfx)}\calB$ into itself, so that the
Lie derivatives are 
\[
\mfL_\bfv \Fe =  - \bfv\dotnabla \Fe +(\nabla\bfv)\Fe \quad \text{and} \quad
\mfL_\bfv \Fp = - \bfv \dotnabla \Fp.
\] 
Inserting these relations into the kinematic relation for $\pl_t\bfF$ we obtain
an additive kinematic relation between the elastic and the plastic strain rates
in the form
\begin{equation}
  \label{eq:KinRel.Fe}
  \Fe^{-1}\big(\pl_t \Fe{+} \mfL_\bfv \Fe\big) + \big(\pl_t \Fp {+} \mfL_\bfv
  \Fp\big) \Fp^{-1} =0,
\end{equation}
which is the counterpart to the often-used split in the Lagrangian setting, see
e.g.\  \cite{Lee69EPDF, BesVan94MMID, GurAna05DMSP}.

We will describe our full model for thermo-elastoplasticity by the state vector
\[
q = \big(\bfpi,\bfF,\Fp,\tau)^\top,
\]
where $\bfpi= \rho\bfv$ is the momentum and $ \tau $ is a scalar-valued
thermodynamical field variable such as the internal energy, enthalpy, entropy,
or temperature. At this stage the structure remains more transparent, when stay
more general, see \cite{Miel11FTDM, BetSch19EMEC}. 

The class of models
considered here can also be described by a smaller set of variables, namely
$(\bfpi,\Fe,\tau)$, see \cite{MieRou25GTMF}, where then \eqref{eq:KinRel.Fe} is
rewritten as $ \pl_t \Fe= -\mfL_\bfv \Fe -\Fe \bfL_\rmp$ and $\bfL_\rmp$ is
modeled accordingly. We hope that the present formulation is clearer, as it
treats $\Fp$ as an integral part of the state of the system.

\subsection{Energy, entropy and driving forces}
\label{su:EnerEntrDriv}

The total energy is given as a the sum of the kinetic energy and the stored
energy in the form
\begin{equation}
  \label{eq:def.calE}
  \calE(\bfpi,\bfF,\Fp,\tau)= \int_\calS\Big(\frac{\det\bfF}{2\rho_\mafo{ref}}
  |\bfpi|^2 + E( \bfF, \Fp, \tau ) \Big)\dd x .
\end{equation}
Here we have written $E$ as a general function of $\bfF$, $\Fp$, and $\tau$,
however, often it is assumed that $E$ depends on $\bfF$ only through $\Fe=\bfF
\Fp^{-1}$ which is a consequence of the Kr\"oner-Lee-Liu decomposition 
\eqref{eq:KronerLeeLiu}. A typical choice would be
$E(\bfF,\Fp,\tau)=W(\bfF\Fp^{-1},\tau) + H(\Fp,\tau)$, where the latter term
can be used to describe (kinematic) hardening effects. At this stage the
structure remains clearer, if we stay with the general function $  E( \bfF,
\Fp, \tau )$, but later on, we will specify to the case  $  E( \bfF,
\Fp, \tau )= \wt  E( \bfF \Fp^{-1} , \Fp, \tau )$. 

Similarly, the entropy is given in the form 
\begin{equation}
  \label{eq:def.calS}
  \calS(\bfF,\Fp,\tau)= \int_\sfS  S(\bfF ,\Fp, \tau ) \dd \sfx ,
\end{equation}
where no dependence on the momentum $\bfpi$ is present because of Galilean
invariance. Here $S$ is assume to be strictly increasing in $\tau$, and the
temperature is defined via 
\[
\theta = \Theta(\bfF,\Fp,\tau): =\frac{\pl_\tau  E(\bfF,\Fp,\tau)}{\pl_\tau 
  S(\bfF,\Fp,\tau)} 
\quad\Longleftrightarrow \quad \tau=\wh \tau(\bfF,\Fp,\theta). 
\]

Note that $E$ and $S$ are an densities with respect to the spatial domain
$\sfS\subset\R^3$ and that they must be frame indifferent, i.e.\
$E(\bfQ\bfF, \Fp,\tau)= E(\bfF,\Fp,\tau)$ and
$S(\bfQ\bfF,\Fp,\tau)= S(\bfF,\Fp,\tau)$ for all $\bfQ\in \mafo{SO}(\R^d)$,
which implies that $ \pl_{\bfF}E(\bfF,\Fp,\tau)\bfF^*$ and
$ \pl_{\bfF}S(\bfF,\Fp,\tau)\bfF^* $ are symmetric. Here $\bfF^*$ is the
adjoint operator to $\bfF:\rmT_\rmx\rmM\to \rmT_\sfx\sfS$ with
$\sfx=\bfy(t,\rmx)$, i.e.\ $\bfF^*:\rmT^*_\sfx\sfS \to \rmT^*_\rmx\rmM$, and we
will also use $\bfF^{-*}=(\bfF^*)^{-1}=(\bfF^{-1})^*:
\rmT^*_\rmx\rmM\to\rmT^*_\sfx\sfS$.\medskip 

As GENERIC structures suggest, the main driving forces for the reversible
(Hamiltonian) dynamics are $\rmD\calE(q)$ and those for the dissipative
dynamics are $\rmD\calS(q)$:
\[
\rmD\calE(q)=\bma{@{}c@{}} \bfv \\ 
   \pl_{\bfF}E(..)+\frac{\rho|\bfv|^2}2\,\bfF^{-*} \\[0.3em]
   \pl_{\Fp} E(..) \\[0.2em]   \pl_\tau  E(..) \ema 
\quad \text{and}\quad  
\rmD\calS(q)= \bma{@{}c@{}}0 \\  \pl_{\bfF}S(..) \\[0.3em]
  \pl_{\Fp} S(..) \\[0.2em]   \pl_\tau  S(..) \ema \!.
\]
However, in light of the
special GENERIC structure introduced at the end of Section \ref{su:SetupGEN},
it will be more natural to look at the combined generalized driving forces
$M_\calS(q)\rmD\calE(q) $ and $M_\calE(q) \rmD\calS(q)$. 

\begin{proposition}
\label{pr:MSDE.MEDS}
With $e_\tau=(0,0,0,1)^\top$ we have the relations 
\begin{equation}
  \label{eq:DrivForces}
 M_\calS(q)^*\rmD\calE(q)=\bma{@{}c@{}} \bfv \\[0.2em] \!
   \bfSigma^F_\rme + \frac{\rho|\bfv|^2}2\,\bfF^{-*} \! \\[0.2em]
  \bfSigma^F_\rmp\\[0.2em] \Theta \ema\!  \text{ and }
 M_\calE(q)^*\rmD\calS(q)=  \tdfrac{-1}{\Theta(q)}
 M_\calE(q)\rmD\calS(q)+\tdfrac{1{+}\Theta}\Theta e_\tau   ,
\end{equation}
where $\bfSigma^F_\rme=  \pl_{\bfF}E(..){-} \Theta(..) \pl_{\bfF}S(..)$ and
$\bfSigma^F_\rmp= \pl_{\Fp}E(..){-} \Theta(..) \pl_{\Fp}S(..) $.  

Here $\bfSigma^F_\rme$ and $\bfSigma^F_\rmp$ are the derivatives of the free energy
$F(q)=E(q){-}\Theta(q)S(q)$ at fixed temperature with respect to $\bfF$ and
$\Fp$, respectively. Inverting the relation $\theta=\Theta(w,\tau)$ into
$ \tau = \wh\tau(w,\theta)$ and setting $\Psi(w,\theta):=F\big(w, \wh \tau
(w,\theta)\big)$, this means 
\[
 \bfSigma^F_\rme= \pl_{\bfF} \Psi(\bfF,\Fp,\theta)\big|_{\theta=\Theta(\bfF,\Fp,\tau)}
 \quad \text{and} \quad 
 \bfSigma^F_\rmp= \pl_{\Fp} \Psi(\bfF,\Fp,\theta)\big|_{\theta=\Theta(\bfF,\Fp,\tau)}.
\]
\end{proposition}
\begin{proof} 
The transformation matrices in Section
\ref{su:SetupGEN} read  
\begin{equation}
  \label{eq:Rel.MS.ME}
  M_\calS(q)^*=I - \tfrac1{\pl_\tau S}\big(\rmD\calS(q) {-}e_\tau\big)\oti e_\tau
\text{ and } 
M_\calE(q)^*=I - \tfrac1{\pl_\tau E}\big(\rmD\calE(q){-}e_\tau\big)\oti e_\tau.
\end{equation}
Thus, the generalized driving forces take the form 
\[
 M_\calS(q)^*\rmD\calE(q)= \rmD\calE(q)- \Theta \rmD\calS(q)  + \Theta e_\tau 
\quad\text{and} \quad 
M_\calE(q)^*\rmD\calS(q) = \rmD\calS(q) - \frac1\Theta \rmD\calE(q) +
\tdfrac1\Theta \;\! e_\tau,  
\]
which means that the two driving forces, except for the last scalar component,
are the same up to constant $- \Theta(..)$. Hence, the second statement in
\eqref{eq:DrivForces} is shown.  

The first relation in \eqref{eq:DrivForces} follows simply by using
\eqref{eq:Rel.MS.ME}, the formulas for $\rmD\calE$ and $\rmD\calS$, and the
definition of $\Theta$. 

The last statement about the free energy is established in
\cite[Eqn.\,(2.13)]{Miel11FTDM}.  
\end{proof} 

In fact, for the dissipative driving forces, it will be more useful to
generalize $M_\calE(q)^*$ to the better adapted operator $N_\calE$. We
emphasize that the conditions \eqref{eq:SpecGEN.MS.ME} for 
$M_\calS$ are much more restrictive, because the third condition asks $M_\calS$
to be the inverse of a derivative. In contrast, the condition  
\eqref{eq:SpecGEN.si.f*g*} for $M_\calE$ are simpler, e.g.\ $M_\calE\equiv 0$
would be allowed (but not really useful). 
In the following we will use the operator 
\begin{equation}
  \label{eq:def.NE*}
  N_\calE(q)^*=\bma{@{}cccc@{}}\bfD(\Box)\!\!&0&0& 
  \tdfrac{-\Box}{\pl_\tau E}\, \bfD(\bfv)\\[0.6em]
 0&0&\!\Box\!&\!\tdfrac{-\Box}{\pl_\tau  E}\, \pl_{\Fp} E 
 \\[0.5em]
0&0&\:0\:& \Box /\pl_\tau  E
\ema\!, \ \ { \text{i.e.\ } N_\calE(q)^* \! 
  \bma{@{}c@{}}\bfw\\ \bfxi\\ \bfeta\\ \kappa \ema 
 = \bma{@{}c@{}} \!\bfD(\bfw){-} 
       \frac{\kappa}{\pl_\tau E}\, \bfD(\bfv)\\[0.6em] 
   \bfeta- \tdfrac{\kappa}{\pl_\tau  E}\, \pl_{\Fp} E
   \\[0.6em] \tdfrac{\kappa}{\pl_\tau  E} \ema\!. }
\end{equation}
By construction, we have $N_\calE(q)^*\rmD\calE(u)=(0,0,1)^\top $ and find
the adapted driving forces  
\begin{equation}
  \label{eq:def.bfeta.ME*DS}
\bfeta=\bma{@{}c@{}}\bfeta_\rmm \\ \bfeta_\rmp\\ \eta_\rmt \ema =  
N_\calE(q)^*\rmD\calS(q) = \bma{@{}c@{}} -\tdfrac1\Theta\, \bfD(\bfv) \\[0.4em]
 \tdfrac1\Theta\big(\pl_{\Fp} S - \Theta \pl_{\Fp} E \big)\\[0.4em]
  1/\Theta   \ema  
= \frac1\Theta \bma{@{}c@{}} -\bfD(\bfv) \\[0.4em] 
- \bfSigma^F_\rmp\\   1  \ema . 
\end{equation}
The major reason for choosing $N_\calE(q)^*$ is that the arising driving forces
are Galilean invariant, in contrast to $M_\calE(q)^*\rmD\calS(q)$ which
involves $\bfv$ and $\rho|\bfv|^2$. Thus, we will have more flexibility in
making physically reasonable choices for $\calR^*_\mafo{simple}$.  Note also
that $N_\calE(q)^*$ is not quadratic and hence not invertible (in contrast to
$M_\calS(q)^*$). This reflects the fact that the equation of $\bfF$ is purely
kinematic and thus will not include any dissipative effects.

\subsection{The Poisson structure for the Hamiltonian part}
\label{su:Jacobi.LeiDer}

The main result of paper states that skew-symmetric operators $\bbJ$ build with
Lie derivatives automatically satisfy the Jacobi identity and hence qualify as
Poisson operators for Eulerian continuum mechanics. 

\begin{theorem}[Jacobi  identity via Lie derivatives]
\label{th:JacobiLie}
We consider the case that the state $\bfz$ is given by $\bfz=(\bfpi,\bfX)$ where
$\bfpi$ is the momentum (considered as a $(d{-}1)$ form as in
\eqref{eq:Momentum}) and a 
collection of variables comprised into a tuple $\bfX$. We define the
state-dependent skew-symmetric operator $\bbJ(\bfpi,\bfX)$ via 
\[
\bbJ(\bfpi,\bfX) \binom{\bfv}{\bfxi} 
= \bma{cc}-\mfL_\Box \bfpi& \bbB(\bfX)\Box \\ -\mfL_\Box \bfX & 0 \ema
\binom{\bfv}{\bfxi}  = \binom{-\mfL_\bfv \bfpi + \bbB(\bfX)\bfxi}{-\mfL_\bfv
  \bfX} ,
\]
where $\bbB(\bfX)$ is defined by skew-symmetry, namely $\langle
\bfv,\bbB(\bfX)\bfxi\rangle = \langle\bfxi, \mfL_\bfv \bfX\rangle$.  

Then, $\bbJ$ satisfies the Jacobi identity \eqref{eq:GEN.Jacobi}, and hence
provides a Poisson structure. 
\end{theorem}
\begin{proof}
The proof relies in the rules for Lie derivatives, in particular the commutator
rule \eqref{eq:LieCommutator}, and the simple observation that the mapping
$(\bfpi,\bfX) \mapsto 
\bbJ(\bfpi,\bfX)$ is linear. A special case of our result was already obtained
in \cite[Prop.\,A.1]{MieRou25GTMF}, however the present proof is significantly
shorter and still self-contained. 

We consider on of the three terms in the Jacobi identity and simplify it
systematically using $\rmD\bbJ(\bfpi,\bfX)[\bfpi_2,\bfX_2]=  \bbJ(\bfpi_2,\bfX_2)$
(because of linearity), where
$(\bfpi_2,\bfX_2)^\top=\bbJ(\bfpi,\bfX)(\bfv_2,\bfxi_2)^\top$. We obtain
\begin{align*}
\mathsf T_{1,2,3}&:=\big\langle \tbinom{\bfv_1}{\bfxi_1},
\rmD\bbJ(\bfpi,\bfX)[\bfpi_2,\bfX_2]\tbinom{\bfv_3}{\bfxi_3}\big\rangle =
 \langle \bfv_1, -\mfL_{\bfv_3}\bfpi_2 {+} \bbB(\bfX_2)\bfxi_3\rangle 
- \langle \bfxi_1,\mfL_{\bfv_3}\bfX_2\rangle
\\
&=\langle \mfL_{\bfv_3}\bfv_1, \bfpi_2\rangle + \langle
\bfxi_3,\mfL_{\bfv_1}\bfX_2\rangle - \langle \bfxi_1,\mfL_{\bfv_3}\bfX_2\rangle
\end{align*}
Inserting the definitions of $\bfpi_2= -\mfL_{\bfv_2}\bfpi+ \bbB(\bfX) \bfxi_2$
and 
$\bfX_2=-\mfL_{\bfv_2}\bfX$ and using the definition of $\bbB(\bfX)$ we obtain 
\begin{align*}
\mathsf T_{1,2,3}&= -\langle \BBL \bfv_3,\bfv_1\BBR ,\mfL_{\bfv_2}\bfpi\rangle 
  + \langle \bfxi_2, \mfL_{\BBL \bfv_3,\bfv_1\BBR }\bfX\rangle  
 - \langle \bfxi_3,\mfL_{\bfv_1}\mfL_{\bfv_2}\bfX\rangle 
 + \langle \bfxi_1,\mfL_{\bfv_3} \mfL_{\bfv_2}\bfX\rangle.
\end{align*}
The first term can be rewritten as
$\langle \mfL_{\bfv_2}\BBL \bfv_3,\bfv_1\BBR ,\bfpi\rangle = \big\langle \big[\!\big[
\bfv_2, \BBL \bfv_3,\bfv_1\BBR  \big]\!\big], \bfpi \big\rangle$. Hence adding the
corresponding cyclic 
permutations, we can exploit the Jacobi identity for vector fields in
\eqref{eq:JacoId.VectFie} and find that the contribution of the terms linear in
$\bfpi$ in $\mathsf C:= \mathsf T_{1,2,3} +\mathsf T_{2,3,1}+\mathsf T_{3,1,2}$
cancel each other.

Similarly, we can look at all terms in $\mathsf C$ involving $\bfxi_1$ and find
\[
\langle \bfxi_1,\mfL_{\bfv_3} \mfL_{\bfv_2}\bfX\rangle 
-\langle \bfxi_1,\mfL_{\bfv_2} \mfL_{\bfv_3}\bfX\rangle 
+ \langle \bfxi_1, \mfL_{\BBL \bfv_2,\bfv_3\BBR }\bfX\rangle  =0
\]
by using the commutator rule \eqref{eq:LieCommutator}. The same holds for the
terms involving $\bfxi_2$ and $\bfxi_3$, and hence the Jacobi identity for
$\bbJ$ is established. 
\end{proof}

In the sense of the special form of GENERIC we define $\bbJ_\mafo{simple}$ by
using the special choice $\tau=s$, namely the entropy density. The reason is
that we know that the Hamiltonian dynamics does not change the entropy. Hence,
we know that the evolution should be the simple transport along the Eulerian
vector field $\bfv$ as a extensive variable. Note also that $\Fp$ is an
intensive variable that is $\bfU$-valued with $\Fp(t,x) \in
U_x=\mafo{GL}(\rmT_x\rmM))$. 
Denoting the Lie derivatives for
$\bfpi$, $\bfF$, $\Fp$, and $s$ by $\mfL_\bfv^\mafo{mo}$,
$\mfL^\mafo{ve}_\bfv$, $\mfL_\bfv^\mafo{in}$, and $\mfL_\bfv^\mafo{ex}$,
respectively, we set
\[
\bbJ_\mafo{simple}(\bfpi,\bfF,\Fp,s) = \bma{cccc} 
-\mfL_\Box^\mafo{mo}\bfpi &\bbB^\mafo{ve}(\bfF)\Box 
   &\bbB^\mafo{in}(\Fp)\Box & \bbB^\mafo{ex}(s)\Box
\\
-\mfL^\mafo{ve}_\Box \bfF  &0&0&0 \\
-\mfL_\Box^\mafo{in} \Fp&0&0&0\\ 
-\mfL_\Box^\mafo{ex} s&0&0&0  \ema
\]
where the operator in the first row are obtained by skew symmetry from the
first column:
\[
\bbB^\mafo{ve}(\bfF)\bfXi_\rme= \nabla \bfF \mdot \bfXi_\rme +
\DIV\!\big(\bfXi_\rme\bfF^*\big) , \quad  
   \bbB^\mafo{in}(\Fp)\bfXi_\rmp = \nabla \Fp \mdot \bfXi_\rmp, \quad 
 \bbB^\mafo{ex}(s)\xi=-s\nabla \xi.
\]

The full Poisson structure $\bbJ(\bfpi,\bfF,\Fp,\tau)$ is now obtain by using
the transformation $s=S(\bfF,\Fp,\tau)$ inducing the transformation operator
$M_\calS$ resulting in 
\[
\bbJ(\bfpi,\bfF,\Fp,\tau)= M_\calS(\bfpi,\bfF,\Fp,\tau) \bbJ_\mafo{simple}
\big(\bfpi,\bfF,\Fp,S(\bfF,\Fp,\tau)\big) M_\calS(\bfpi,\bfF,\Fp,\tau)^*.
\]
From Theorem \ref{th:JacobiLie} we know that $\bbJ_\mafo{simple}$ is a Poisson
structure, hence the transformation shows that $ \bbJ$ is again a Poisson
structure.

\section{The Eulerian form of thermo-visco-elastoplasticity}
\label{se:EulThermoVisco}

We now collect the evolutionary equations for thermo-visco-elastoplasticity at
finite strain from the GENERIC form. Abbreviating $q=(\bfpi,\bfF,\Fp,\tau)$ we
have 
\begin{align*}
&\pl_t q = \bfV_\mafo{Ham}(q) + \bfV_\mafo{diss}(q) \quad \\
&\text{with} \quad 
 \bfV_\mafo{Ham}(q)= \bbJ(q)\rmD\calE(q) \ \text{ and } \ 
 \bfV_\mafo{diss}(q)= N_\calE(q) \pl_\xi \calR^*_\mafo{simple} \big( q,
 N_\calE(q)^*\rmD\calS(q) \big).
\end{align*}
One of the advantages of the GENERIC framework is that we can derive the 
Hamiltonian part and the dissipative part completely independently, thus
separating the two quite different physical phenomena.

\subsection{The Hamiltonian part of the model}
\label{su:HamiltonianPart}

We now discuss the terms arising from of the Hamiltonian part of the dynamics,
namely $\plt q= \bfV_\text{Ham}(q) = \bbJ(q)\rmD\calE(q)$.  Using the form of
$\calE$ and the definition of $\bbJ$ 
we obtain 
\begin{equation}
  \label{eq:HamPart}
\bfV_\text{Ham}(q)=\bbJ(q)\rmD\calE(q)  = \bma{c} 
 -\DIV\big(\rho\bfv{\otimes}\bfv\big) + \DIV \bfSigma_\text{Cauchy}\\[.25em]
  -\bfv\dotnabla  \bfF + (\nabla\bfv)\,\bfF \\[.25em]
     -\bfv\dotnabla  \Fp \\[.05em]
  \! - \bfv\dotnabla  \tau 
  - \tdfrac{S(..)\DIV\bfv  + \pl_\bfF S(..)\bfF^* \mdot  \bfD(\bfv)}
    {\pl_\tau S(..)} \!\ema\! ,
\end{equation}
where the Cauchy stress tensor $\bfSigma_\text{Cauchy}$ is defined via the
free-energy density $\Psi=E-\Theta S$:
\[
\bfSigma_\text{Cauchy} = \bfSigma^F_\rme  \bfF^*+ \Psi\,\bbI
\quad \text{with }\  \bfSigma^F_\rme= \pl_\bfF E(\bfF,\Fp,\tau)-\Theta(\bfF,\Fp,
\tau) \;\! \pl_\bfF S(\bfF,\Fp ,\tau). 
\] 
A simplified version of \eqref{eq:HamPart} was derived in 
in \cite[Sec.\,3]{HutTer08FAEM}, where $\Fp$ is neglected and $tau=\theta$ is
chosen.

The rest of this subsection will explain the form of the four components  of
$ \bfV_\text{Ham}(q)$   in detail. 
For this it is important to keep the adapted reversible driving forces 
$M_\calS(q)^*\rmD\calE(q)$  from \eqref{eq:DrivForces} in mind, where the first
component remains $\bfv$. This vector will be applied to
$M_\calS(q)\bbJ_\mafo{simple}(\Phi(q))$ to yield $\bfV_\text{Ham}(q)=$ 
{\renewcommand{\arraycolsep}{0.3em}%
\[
\bma{cccc}I&0&0&0\\ 0&I&0&0\\0&0&I&0\\
 -\tfrac{\pl_\bfv S}{\pl_\tau S}&  -\tfrac{\pl_\bfF S}{\pl_\tau S}& 
 -\tfrac{\pl_{\Fp} S}{\pl_\tau S}& 
 \tfrac{1}{\pl_\tau S}  \ema 
\bma{cccc}-\mfL_\Box^\mafo{mo}\bfpi &\bbB^\mafo{ve}(\bfF)
   &\bbB^\mafo{in}(\Fp) & \bbB^\mafo{ex}(S)
\\
-\mfL^\mafo{ve}_\Box \bfF  &0&0&0 \\
-\mfL_\Box^\mafo{in} \Fp&0&0&0\\ 
-\mfL_\Box^\mafo{ex} S&0&0&0  \ema 
\bma{c}\bfv\\[0.3em] 
 \!\bfSigma^F_\rme{+} \tfrac{\rho|\bfv|^2}2 \bfF^{-*}\! 
\\[0.3em]
\bfSigma^F_\rmp\\ \Theta \ema 
\]
}%
Note that the first matrix $M_\calS(q)$ does not change
the first three components; hence, the second and third
components of $ \bfV_\text{Ham}(q)$ are simply given by $- \mfL^\mafo{vec}_\bfv
\bfF$ and $- \mfL^\mafo{int}_\bfv \Fp$. 

For the fourth component we can take advantage from $pl\bfv S\equiv 0$; hence
it is a linear combination of the three Lie derivatives, namely 
\[
\frac1{\pl_\tau S} \Big(\pl_\bfF S  \mdot \mfL^\mafo{ve}_\bfv \bfF + \pl_{\Fp}  S
   \mdot  \mfL_\bfv^\mafo{in} \Fp - \mfL_\bfv^\mafo{ex} S \Big) 
  = - \bfv\dotnabla  \tau 
  -\frac{ S(..)\DIV\bfv  + \pl_\bfF S(..)\bfF^* \mdot  \bfD(\bfv)}{\pl_\tau S(..)}. 
\]
For the last identity, we use that the terms involving $\bfv\dotnabla \bfF$
and $\bfv\dotnabla \Fp$ arising from $\mfL^\mafo{ve}_\bfv \bfF$ and $
\mfL_\bfv^\mafo{in} \Fp$, respectively, cancel with those arising from $
\mfL_\bfv^\mafo{ex} S = \DIV(S(..)\bfv)$. The last term in the above relation
equals  $ \pl_\bfF S(..)\mdot\big((\nabla\bfv)\bfF^*\big)$ because of the symmetry
of $  \pl_\bfF S(..)\bfF^*$. 

Thus, it remains to establish the simple representation of the first component
of $\bfV_\text{Ham}(q)$ in \eqref{eq:HamPart}. For this, we first calculate the
terms involving $\bfv$  (see also \cite[Eqn.\,(4.6)]{ZaPeTh23GFRF}), namely 
\begin{align*}
&-\mfL_\bfv \bfpi + \bbB^\mafo{ve}(\bfF)\tfrac{\rho|\bfv|^2}2 \bfF^{-*}
= - \DIV(\bfpi{\otimes}\bfv) 
 {-}  (\nabla\bfv)^*\bfpi 
 {+} \frac{\rho|\bfv|^2\!\!}2 \bfF^{-*}  \mdot  \nabla \bfF 
 {+} \DIV\Big(\frac{\rho|\bfv|^2\!\!}2\,\bbI\Big) \\
&\quad =  - \DIV(\rho\bfv{\otimes}\bfv) 
  - \rho \,  \nabla \big(\tfrac{|\bfv|^2}2\big) 
   + \tfrac{\rho|\bfv|^2}{2\det\bfF} \nabla \det\bfF 
   +  \nabla\big( \rho \tfrac{|\bfv|^2}2 \big) 
 =  - \DIV(\rho\bfv \oti \bfv),
\end{align*}
because $\rho=\rho_\mafo{ref}/\det\bfF$ implies
$\nabla\rho= -(\rho/\det\bfF)\nabla\det\bfF$, where $\rho_\mafo{ref}$ is a
constant.  

To simplify the the remaining terms in the first component we first use
$\Psi=E-\Theta S$ to obtain
\begin{align*}
\nabla \Psi &= \nabla E- \Theta \nabla S - S \nabla \Theta \\
&= 
 \nabla\bfF\mdot (\pl_\bfF E{-}\Theta \pl_\bfF S) + \nabla\Fp \mdot 
(\pl_{\Fp} E{-}\Theta \pl_{\Fp} S) +
\underbrace{(\pl_\tau E{-}\Theta \pl_\tau S)}_{=0} \nabla \tau -S \nabla \Theta 
\\
&= 
 \nabla\bfF\mdot \bfSigma^F_\rme + \nabla \Fp \mdot \bfSigma^F_\rmp -S \nabla \Theta ,
\end{align*}
where we recall $\bfSigma^F_{\rme,\rmp}$ from Proposition
\ref{pr:MSDE.MEDS}. Hence, the remain terms are
\begin{align*}
&\bbB^\mafo{ve}(\bfF)\bfSigma^F_\rme + \bbB^\mafo{in}(\Fp)\bfSigma^F_\rmp +
\bbB^\mafo{ex}(S) \Theta \\
&= \DIV(\bfSigma^F_\rme \bfF^*) + \nabla\bfF\mdot \bfSigma^F_\rme + \nabla \Fp
\mdot \bfSigma^F_\rmp -S \nabla \Theta  = \DIV\big(\bfSigma^F_\rme \bfF^* + \Psi I\big).
\end{align*}
Hence, also the desired form of the first component in \eqref{eq:HamPart} is
established.

\subsection{The dissipative part of Eulerian thermo-elastoplasticity}
\label{su:GEN.diss}

The modeling of the dissipative effects is considerably simpler than that of
the reversible Hamiltonian part, because the their is much more freedom in
choosing dissipation potentials than in choosing Poisson structures. One of the
main advantages of GENERIC is indeed the very structured modeling of
dissipative processes allowing for general couplings. Using the subdifferential
of the dual dissipation potential will automatically satisfy the so-called
Onsager symmetries. In the quadratic case $\calR^*(q,\xi)=\frac12\langle
\xi, \bbA(q)\xi\rangle $ this is seen by the fact that $\pl_\xi \calR^*(q,\xi)
= \bbK(q) \xi$ with $\bbK(q)=\frac12(\bbA(q){+}\bbA(q)\big)= \bbK(q)^*$, which
is Onsager's reciprocal relation. Note that $\bbK(q)=\rmD^2\calR^*(q,0)$ is the
symmetric Hessian of $\calR^*(q,\cdot)$. 

Using the special form of GENERIC as described in \eqref{eq:SpecGen.all} 
we can construct suitable nonlinear dissipation potentials $\calR^*$
(or linear Onsager operators $\bbK$) by collecting the building
blocks of the dissipative effects and combining them
with a nontrivial operator $N_\calE$ in the form
\[
\calR^*(q,\bfzeta)=\calR^*_\mathrm{simple}(q,N_\calE(q)^*\bfzeta) \quad
\text{or} \quad \bbK(q)=N_\calE(q)\bbK_\mathrm{simple}(q)N_\calE(q)^*.
\]
This strategy is propagated in \cite[Sec.\,2.3.2]{Otti05BET} and follows
\cite{Edwa98ASDG}. 

In our model we can have three dissipative processes:

(A) viscoelastic dissipation induced by $\bfD(\bfv)=\frac12\big(\nabla
\bfv{+}(\nabla\bfv)^\top \big)$,

(B) plastic dissipation induced by $\pl_t\Fp$, 

(C)  heat flow induced by $\nabla (1/\theta)$. 

\noindent
Thus, we can construct a suitable dual dissipation potential in the additive form 
\[
 \calR^*_\mathrm{simple} = \calR^*_\rmA + \calR^*_\rmB + \calR^*_\rmC. 
\]
However, we emphasize that this simplistic assumption is by far not
necessary. Of course, it is possible to construct much more general
thermodynamically consistent models where there is a strong interaction
of the different dissipation mechanics, but to keep simplicity and clarity we
will restrict our approach to the case of a simple block structure.\smallskip

The advantage of using an operator $N_\calE$ is three-fold. First, it is used 
to guarantee the second non-interaction condition by asking
\begin{equation}
  \label{eq:NE.2ndNIC}
    N_\calE(q)^* \rmD\calE(q)=e_\tau=(0,...,0,1)^\top \quad \text{and} \quad
  \calR^*_\mathrm{simple}\big(q,\lambda\;\!e_\tau)^\top\big)=0\ \text{ for all
  }\  \lambda \in \R. 
\end{equation}
Here $\lambda\in \R$ stand for the constant (reciprocal of the) 
temperature $1/\theta$, which does not generate any dissipation.

The second advantage is the fundamental observation in
\cite[Sec.\,4.3]{Miel11FTDM} that the dissipative
(a.k.a.\ irreversible) driving forces are now given by the generalized driving
forces 
\[
\bfeta= N_\calE(q)^*\rmD\calS(q)\,,
\]
which contains important thermodynamical information by identifying the
correct combinations of $\rmD\calS$ and $\rmD \calE$ governing dissipative
processes and satisfying Galilean invariance.

Finally, the operator $N_\calE$ acting from the left on
$\pl\calR^*_\mathrm{simple}$ adjusts the dissipative terms in such a way that 
energy conservation holds for 
\[
  \pl_\xi \calR^*(q,\rmD \calS(q)) = N_\calE(q)\,
 \pl\calR^*_\mathrm{simple} \big( q, N_\calE(q)^*\rmD\calS(q) \big). 
\]
Moreover, this form involving $N_\calE$ and $N_\calE^*$ guarantees the Onsager
symmetries: if $\calR^*_\mafo{simple}(q,\bfeta)$  $=\frac12\langle \bfeta,
\bbK_\mafo{simple}(q)\bfeta\rangle$ is quadratic, then $\calR^*$ is still
quadratic with $\calR^*(q,\bfxi)=\frac12\langle \bfxi,
\bbK(q)\bfxi\rangle$, where $\bbK(q)=N_\calE(q)\bbK_\mafo{simple}(q)
N_\calE(q)^* = \bbK(q)^*$ is again symmetric.   

We recall our special choice for $N_\calE(q)^*$ from \eqref{eq:def.NE*} and
find the adjoint
\[
N_\calE(q)=\bma{ccc} -\DIV(\Box)&0&0\\
 0 &0&0\\ 0&I&0\\
\tdfrac{-1}{\pl_\tau E} \bfD(\bfv)\mdot\Box& 
\tdfrac{-1}{\pl_\tau E}\pl_{\Fp}E \mdot \Box&\tdfrac1{\pl_\tau E} \ema .
\]
The three $0$'s in the second row indicate that dissipative forces cannot
change the kinematic relation of $\bfF$. 

The full dual dissipation potential takes the form
\[
\calR^*(q,\bfzeta)=\calR^*_\mathrm{simple}(q,N_\calE(q)^*\bfzeta)\,,
\]
and now assume that $\calR^*_\mathrm{simple}$ has a block structure
\begin{align*}
\calR^*_\mafo{simple}(q,\bfeta) &=\calR^*_\mafo{vi.el}(q,\bfeta_\rmm) +
\calR^*_\mafo{vi.pl} (q,\bfeta_\rmp)+  \calR^*_\mafo{heat}(q,\eta_\rmt).
\end{align*}
Yet, we hasten to say that this is a simplification that is not necessary at
all; in fact, it is one of the big advantages of the GENERIC framework
that it can easily handle coupling phenomena between different effects. 

Even for the these five blocks we only write the simplest forms and leave the
study of more general dissipation potentials to future work. 
\begin{subequations}
 \label{dissip-pot}
 \begin{align}
  \label{dissip-pot-a}
  &\calR^*_\mafo{vi.el}(q,\bfeta_\rmm) = \int_\sfS \frac{\Theta}2 
   \bfeta_\rmm \mdot \bbD_\mafo{vi.el} (q)\bfeta_\rmm\dd \sfx,
   \quad
   \calR^*_\mafo{vi.pl} (q,\bfeta_\rmp) = \int_\sfS\! R_\mathrm{vi.pl}^*
   \big(q,\Theta \Fp^*\bfeta_\rmp \big)\dd\sfx,
  \\
  &\calR^*_\mafo{heat}(q,\eta_\rmt) =\int_\sfS \frac12 \nabla
   \eta_\rmt {\cdot}\bbK_\mafo{heat}(q) \nabla \eta_\rmt\dd\sfx .
 \end{align}
\end{subequations}
Here the pre-multiplication of $\bfeta_\rmp$ by $\Fp^*$ will generate the
so-called plastic indifference, see \cite[Sec.\,3.1]{Miel03EFME} and below.  

With these choices the dissipative (irreversible) part
$\bfV_\mathrm{diss}(q) = \pl_\bfzeta \calR^*\big(q,\rmD\calS(q)\big)$ of the
evolution takes the form
\begin{align*}
\bfV_\mathrm{diss}(q)&= N_\calE(q)\pl_\bfeta \calR^*_\mafo{simple} \big( q,
N_\calE(q)^*\rmD\calS(q)\big) = \bma{@{}c@{}} -\DIV\!\big(
\pl_{\bfeta_\rmm} \calR^*_\mafo{vi.el}(q,\bfeta_\rmm )\big) \\ 0 \\ 
\pl_{\bfeta_\rmp}\calR^*_\mafo{vi.pl}(q, \bfeta_\rmp) \\
\tdfrac1{\pl_\tau E}\, j_\mafo{ener} \ema  
\\& \text{ with } j_\mafo{ener}= -\pl_{\bfeta_\rmm}
\calR^*_\mafo{vi.el}(..)\mdot \bfD (\bfv)
-\pl_{\bfeta_\rmp} \calR^*_\mafo{vi.pl}(..)\mdot \pl_{\Fp}E(q) 
 + \pl_{\eta_\rmt} R^*_\text{heat}(q,\eta_\rmt) , 
\end{align*}
Inserting $\bfeta = N_\calE(q)^*\rmD\calS(q)$ from \eqref{eq:def.bfeta.ME*DS}
and the choices \eqref{dissip-pot} we arrive at 
\begin{align*}
\bfV_\mafo{diss}(q)&=
\bma{c}\DIV\big( \bbD_\mathrm{visc}(q)\bfD \big) \\[.3em]  0
\\[.3em]
\Fp \bfL_\mafo{vi.pl} 
\\[.5em]  
\!\!
 \tdfrac1{\pl_\tau E}\Big( \bfD \mdot  \bbD_\mathrm{visc} \bfD
 +  (\Fp \bfL_\mafo{vi.pl}) \mdot  \pl_{\Fp}E(q)
 - \DIV\!\big(\bbK_\mafo{heat}\nabla\tdfrac1\Theta\big)\Big) 
\!\!   \ema ,
\end{align*}
where $\bfD=\bfD(\bfv)=\frac12\big(\bbI_\mafo{Riesz}\nabla\bfv{+}(\nabla\bfv)^*
\bbI_\mafo{Riesz}\big) $ (cf.\
\eqref{eq:def.bfD.bfv}) and $\bfL_\mafo{vi.pl}= \Theta(q) \pl_{\bfxi_\rmp}
R^*_\mafo{vi.pl}\big(q, \!-\Fp^* \bfSigma^F_\rmp \big) $ with 
$\bfSigma^F_\rmp= \pl_{\Fp} E(q)-\Theta(q) \pl_{\Fp} S(q)$.

\subsection{The Eulerian model for thermo-visco-elastoplasticity}
\label{su:FullGEE}

We can now assemble the whole continuum model for non-isothermal finite-strain
visco-elastoplasticity as obtained from the GENERIC framework:
\[
\dot q = \bfV_\mathrm{Ham}(q) + \bfV_\mathrm{diss}(q)= \bbJ(q)\rmD\calE(q) +
\pl_\zeta \calR^*\big( q, \rmD\calS(q)\big). 
\]
Combining the results from Sections \ref{su:HamiltonianPart} and
\ref{su:GEN.diss}, this leads to the following system for $(\bfv,\sfq)$ with
$\sfq=(\bfF,\Fp,\tau) $:
\begin{subequations}
 \label{system}
  \begin{align}
  \label{system-a}
  \pl_t(\rho\bfv) &= -\DIV\big( \rho \bfv \oti \bfv)
    + \DIV\big(\bfSigma_\mafo{Cauchy}(\sfq) + \bbD_\mathrm{visc}(\sfq) \bfD \big),
  \\[0.3em] 
  \label{system-b}
  \pl_t \bfF &=-\mfL_\bfv \bfF=-(\bfv\dotnabla )\bfF + (\nabla \bfv) \bfF,
  \\[0.3em] 
  \label{system-c}
  \pl_t\Fp &=-\bfv\dotnabla \Fp + \Fp \bfL_\mafo{vi.pl} (\sfq),
  \\[0.3em] \label{system-d}
  \pl_t \tau&=j^S_\mafo{Ham}(\sfq) +j^E_\mafo{diss}(\sfq) -
\tdfrac{1}{\pl_\tau E(\sfq)} \DIV\!\Big( 
  \bbK_\mathrm{heat}(\sfq) \nabla\tdfrac1{\Theta(\sfq)}\Big), 
  \end{align}
\end{subequations}
where $\rho= \rho_\mafo{ref}/\det\bfF$ and where we have 
\begin{subequations}
 \label{system-data}
  \begin{align}
  \label{system-data-a}
  \bfSigma_\mathrm{Cauchy}(\sfq)
   &=\big( \pl_\bfF E(\sfq) {-} \Theta(\sfq) \pl_\bfF S(\sfq) \big)\bfF^*
                        + \big(E(\sfq){-}\Theta(\sfq) S(\sfq) \big) \bbI,
  \\[0.3em] \label{system-data-b}
  \bfL_\mafo{vi.pl}(\sfq)
  &= \Theta(\sfq) \pl_{\bfxi_\rmp} R^*_\mafo{vi.pl} \big( 
     \sfq, \!-\Fp^* \bfSigma^F_\rmp \big)  \text{ with }
     \bfSigma^F_\rmp= \pl_{\Fp} E(\sfq) {-}\Theta(\sfq) \pl_{\Fp} S(\sfq),
  \\[0.3em] \label{system-data-d}
  j^S_\mathrm{Ham}(\sfq)&= - \bfv\dotnabla  \tau   - \tdfrac1{\pl_\tau S(\sfq)}
\Big(S(\sfq)\DIV\bfv  + \pl_\bfF S(\sfq)\bfF^* \mdot  \bfD(\bfv) \Big),
\\[0.3em] \label{system-data-e}
j^E_\mathrm{diss}(\sfq)&=  \tdfrac1{\pl_\tau E(\sfq)} 
  \Big( \bfD(\bfv)\mdot  \bbD_\mathrm{visc}(\sfq) \bfD(\bfv)
 +  \big(\Fp \bfL_\mafo{vi.pl}(\sfq)\big) \mdot  \pl_{\Fp}E(\sfq) \Big).
 \end{align}
\end{subequations}

Specifying the choice of $\tau$ either to $\tau = e$, the density of the
internal  energy  giving $E(\bfF,\Fp,e)=e$, or to $\tau =s$, the density of the
entropy giving $S(\bfF,\Fp,s)=s$, reveals more structure to the terms
$j^S_\mafo{Ham}(\sfq) +j^E_\mafo{diss}(\sfq)$. In the first case we can read of
the heating contribution through mechanical processes, and in the second case
we see the entropy production through plasticity. We refer to
\cite[Sec.\,4.6]{MieRou25GTMF} for a more elaborate discussion.

\section{Discussion and outlook}
\label{se:Conclusion}

In this paper we have brought together the theory of Lie derivatives and the
GENERIC framework to provide the right modeling tools for Eulerian mechanics of
solid materials. 

The GENERIC framework takes care that the model is
thermodynamically consistent in the sense that energy is conserved (first law
of thermodynamics) and that the entropy is non-decreasing (second law of
thermodynamics). 
However, the GENERIC framework enforces an additional structural condition on
the reversible/conservative part and on the irreversible/dissipative part of
the system.  The reversible part has to be strictly Hamiltonian which
is more restrictive than being energy and entropy preserving. For this it would
be sufficient to have the skew-symmetry $\bbJ(q)^*=- \bbJ(q)$ and the
non-interaction condition $\bbJ(q)\rmD\calS(q)\equiv 0$, but GENERIC 
imposes the Jacobi identity \eqref{eq:GEN.Jacobi}. The irreversible part is 
restricted by enforcing that the kinetic relations for the dissipative
processes derive from the (sub)-differential of a dissipation potential, which 
automatically enforces Onsager's symmetry relations. 

Thus, GENERIC systems form a subclass of all thermodynamically consistent
systems. The additional structural conditions may be restrictive but appear to
be valid for many continuum models. In particular, in situations where complex
coupled models are to be developed, the framework provides easy ways for the
derivation of good coupling terms between different variables; see e.g.\
\cite[Sec.\,3.4+4.6]{KMMR19MMSQ}, where a reaction-diffusion system for charge
carriers is coupled to a quantum mechanical system. We expect that the GENERIC
framework will also be very useful in solid mechanics when in addition to
deformation and temperature also internal variables such as a plastic
distortion, phase indicators, dislocation densities, or magnetization have to
be modeled, see e.g.\ \cite{GKRHS16TMFI,BNSS24?VAMC}.  

This approach was used already very successfully in
\cite{Miel11FTDM,HutSve12TMFV,GKRHS16TMFI} for models based on the Lagrangian
description, where a referential body $\rmM$ is used for describing all the
fields. While the Eulerian perspective  working in the spatial domain
$\sfS\subset \R^d$ is commonly used in fluid mechanics, see
e.g.\ \cite{GrmOtt97DTCF12,Otti05BET,ZaPeTh23GFRF}, the combination of 
GENERIC, Eulerian description, and solid mechanics was brought together only
recently in \cite{MieRou25GTMF}. One of the reasons for using the Eulerian
description for solids is the growing mathematical interest in geophysical
flows in the upper mantle of our planet Earth, see e.g.\
\cite{RouTom21CMPE,Roub23TMPE,MiRoSt23?MGDC}. Depending on the time scale under
consideration the solid rock behaves visco-elastically or visco-plastically,
but there are many other variables to be taken into account, like concentrations
of different chemicals, water contents, temperature, or aging variables, see
e.g.\ \cite{GerYue07RCMM,HeGeVa18IRSDF,SBGNW21KFEO}. 

The Eulerian description of solids is intrinsically linked to the notion of Lie
derivatives, because the variables are transported with the velocity $\bfv$ of
the points in the moving and deforming body. This observation from
\cite{MieRou25GTMF} served as a motivation to provide a short
and self-contained introduction to Lie derivatives with a focus on the
applications in continuum thermodynamics in the present paper. 
As an major outcome we found Theorem \ref{th:JacobiLie}, which 
states that in the Eulerian setting the Jacobi
identity \ref{eq:GEN.Jacobi} for Poisson operators $\bbJ$ is intimately
related to the commutator relation \eqref{eq:LieCommutator} for Lie
derivatives. The result is now more general and the proof is shorter and more
direct as for \cite[Thm.\,4.1]{MieRou25GTMF}.

\appendix

\section{Pullbacks and push-forwards in differential geometry}
\label{se:FormDiffGeo}

A diffeomorphism $\bfPhi:M\to N$ defines pull-backs $\bfPhi^*$ and
push-forwards $\bfPhi_*$ as follows:

0. Functions: 
\[
\big(\bfPhi^* g\big)(m)=g(\bfPhi(m)) \qandq \big(\bfPhi_*f\big) (n) = f(\bfPhi^{-1}(n)).
\]

1. Vector fields: 
\begin{align*}
&\big(\bfPhi^*\bfw\big)(m) =\big(\rmD \bfPhi(\bfPhi(m)) \big)^{-1}
\bfw(\bfPhi(m))= \big(\rmD\bfPhi(m)\big)^{-1}\bfw(\bfPhi(m)) \in \rmT_m M 
\\
&\big(\bfPhi_*\bfv\big)(n) =\rmD \bfPhi(\bfPhi^{-1}(n))
\bfv(\bfPhi^{-1}(n)) = \rmD(\bfPhi^{-1})(n) \bfv(\bfPhi^{-1}(n)) \in \rmT_n N
\end{align*}

1$^*$. one-forms: 
\begin{align*}
  &\big(\bfPhi^*\bfbeta\big)(m) =\rmD \bfPhi(m)^*
  \bfbeta(\bfPhi(m)) \in \rmT_m^* M
  \\
  &\big(\bfPhi_*\bfalpha\big)(n) =\big(\rmD \bfPhi(\bfPhi^{-1}(m))\big)^{-*} 
  \bfalpha(\bfPhi^{-1}(n)) 
     \in \rmT_n N
\end{align*}
Here for a matrix $H$ we denote by $H^{-*}$ the transposed of the inverse,
namely $H^{-*}=\big(H^{-1}\big)^*= (H^*)^{-1}$.
Pull-backs and push-forwards are consistent with dual pairings, namely 
\begin{align*}
&\bfPhi^*\big(\langle\bfbeta,\bfw\rangle_N\big)(m)
 = \big\langle \bfPhi^*\bfbeta(m),\bfPhi^*\bfw(m) \big\rangle_{\rmT_mM} 
\\
&\bfPhi_*\big(\langle\bfalpha,\bfv\rangle_M\big)(n)
 = \big\langle \bfPhi_*\bfalpha(\phi^{-1}(n)),\bfPhi_*\bfv(\phi^{-1}(n)) 
    \big\rangle_{\rmT_n N} ,
\\
& \langle \bfPhi^*\bfbeta, \bfv\rangle_M = \langle \bfbeta,
\bfPhi_*\bfv\rangle_N \qandq  \langle \bfalpha, \bfPhi^*\bfw\rangle_M
 = \langle \bfPhi_*\bfalpha, \bfw\rangle_N.
\end{align*}

General tensors $\bfA\in \mfT_{j_\circ}^{i_\circ}(M)$ with $\bfA(m)\in
\mafo{ML}\big( (\rmT_mM)^{i_\circ}\ti (\rmT_mM)^{j_\circ}\big)$\\
and $\bfB\in \mfT_{j_\circ}^{i_\circ}(N)$ with $\bfB(n)\in
\mafo{ML}\big( (\rmT_nN)^{i_\circ}\ti (\rmT_nN)^{j_\circ}\big)$:
\begin{align*}
&\bfPhi^*\bfB(m)[\bfv_1,..,\bfv_{i_\circ},\bfalpha_1,..,\bfalpha_{j_0}]=
\bfB(n) \big[\bfPhi_*\bfv_1,..,\bfPhi_*\bfv_{i_\circ}, 
    \bfPhi_*\bfalpha_1,..,\bfPhi_*\bfalpha_{j_\circ} \big] \Big|_{n=\bfPhi(m)},  
\\
&\bfPhi_*\bfA(n)[\bfw_1,..,\bfw_{i_\circ},\bfbeta_1,..,\bfbeta_{j_0}]=
\bfA(m) \big[\bfPhi^*\bfw_1,..,\bfPhi^*\bfw_{i_\circ},
\bfPhi^*\bfbeta_1,..,\bfPhi^*\bfbeta_{j_\circ} \big]\Big|_{m=\bfPhi^{-1}(n)} .
\end{align*}

\paragraph*{Acknowledgments.}
This research has been partially funded by Deutsche
Forschungsgemeinschaft (DFG) through the Berlin Mathematics Research Center
MATH+ (EXC-2046/1, project ID: 390685689) subproject ``DistFell''. The author
is grateful for stimulating discussions with Tom\'a\v s Roub\'\i\v cek.

\footnotesize

\newcommand{\etalchar}[1]{$^{#1}$}
\def\cprime{$'$}
\providecommand{\bysame}{\leavevmode\hbox to3em{\hrulefill}\thinspace}
\providecommand{\MR}{}

\end{document}